# Automatic Mitigation of Dynamic Atmospheric Turbulence Using Optical Phase Conjugation for Coherent Free-Space Optical Communications

Huibin Zhou[1,*], Xinzhou Su[1], Yuxiang Duan[1], Yue Zuo[1], Zile Jiang[1], Muralekrishnan Ramakrishnan[1], Jan Tepper[2], Volker Ziegler[2], Robert W. Boyd[3], Moshe Tur[4], and Alan E. Willner[1,*]

1. Department of Electrical and Computer Engineering, Univ. of Southern California, Los Angeles, CA 90089, USA
2. Central Research and Technology, Airbus, Taufkirchen, Germany
3. Institute of Optics, University of Rochester, Rochester, New York 14627, USA
4. School of Electrical Engineering, Tel Aviv University, Ramat Aviv 69978, Israel

Corresponding emails*: H.Z. (huibinzh@usc.edu) or A.E.W. (willner@usc.edu)

**Abstracts**

Coherent detection can provide enhanced receiver sensitivity and spectral efficiency in free-space optical (FSO) communications. However, turbulence can cause modal power coupling effects on a Gaussian data beam and significantly degrade the mixing efficiency between the data beam and a Gaussian local oscillator (LO) in the coherent detector. Optical phase conjugation (OPC) in a photorefractive crystal can "automatically" mitigate turbulence by: (a) recording a back-propagated turbulence-distorted probe beam, and (b) creating a phase-conjugate beam that has the inverse phase distortion of the medium as the transmitted data beam. However, previously reported crystal-based OPC approaches for FSO links have demonstrated either: (i) a relatively fast response time of 35 ms but at a relatively low data rate (e.g., <1 Mbit/s), or (ii) a relatively high data rate of 2-Gbit/s but at a slow response time (e.g., >60 s). Here, we report an OPC approach for the automatic mitigation of dynamic turbulence that enables both a high data rate (8 Gbit/s) data beam and a rapid (<5 ms) response time. For a similar data rate, this represents a 10,000-fold faster response time than previous reports, thereby enabling mitigation for dynamic effects. In our approach, the transmitted pre-distorted phase-conjugate data beam is generated by four-wave mixing in a GaAs crystal of three input beams: a turbulence-distorted probe beam, a Gaussian reference beam regenerated from the probe beam, and a Gaussian data beam carrying a high-speed data channel. We experimentally demonstrate our approach in an 8-Gbit/s quadrature-phase-shift-keying coherent FSO link through emulated dynamic turbulence. Our results show ~10-dB improvement in the mixing efficiency of the LO with the data beam under dynamic turbulence with a bandwidth of up to ~260 Hz (Greenwood frequency). Our approach has the potential to significantly increase the resilience of high-performance coherent FSO links to turbulence.

# Introduction

Free-space optical (FSO) communication systems have gained increasing interest in many diverse applications due to the promise of a higher data rate and a lower probability of intercept compared to radio-frequency communications[1–3]. Currently, many FSO link demonstrations use intensity modulation and direct detection (IM/DD)[4]. However, coherent detection with a local oscillator (LO) enables significant and compelling advantages for FSO links, including: (a) better receiver sensitivity, and (b) higher spectral efficiency when utilizing higher-order modulation formats (e.g., quadrature-phase-shift-keying (QPSK) and quadrature-amplitude-modulation (QAM))[5–7].

Unfortunately, atmospheric turbulence is a key challenge in coherent FSO links[6,8]. In a typical coherent detector, a Gaussian data beam efficiently mixes in the photodiode with an LO that has a similar Gaussian modal structure[5]. However, dynamic turbulence (e.g., at the Greenwood frequency of a few hundreds of Hertz[9,10]) can cause wavefront distortion of the transmitted data beam and produce modal power coupling from the transmitted spatial mode (typically, a fundamental Gaussian mode) to many higher-order modes[9,11,12]. Therefore, optoelectronic mixing between the turbulence-distorted multi-mode data beam and a single-Gaussian-mode LO beam becomes significantly degraded in coherent detection (e.g., with mixing loss >20 dB)[12,13]. Various approaches for adaptive dynamic turbulence mitigation in coherent FSO links include (a) adaptive optics by measuring the wavefront distortion and correcting it through an electronic feedback loop[7,14–16], and (b) coherent multi-mode combining by collecting multiple modes and combining them using additional electronic iterative digital signal processing (DSP)[17–22].

Alternatively, it might be highly advantageous to "automatically" mitigate and adapt to dynamic turbulence without the need for electronic signal processing[23–25]. One approach is to use optical phase conjugation (OPC) based on the photorefractive effects in a crystal[26–31], including the following: (i) a probe beam reverse-propagates from the receiver (Rx) to the transmitter (Tx) and experiences distortion due to turbulence; (ii) this probe beam "writes" its turbulence-induced phase distortion into a crystal, (iii) a forward-propagating Tx beam "reads" the crystal and takes on the conjugate of the phase distortion (i.e., the inverse effect) of the turbulence-affected probe beam, and (iv) this conjugate beam propagates through the same turbulence to the Rx and the turbulence distortion is automatically mitigated[29]. Crystal-based OPC was shown using (i) self-pumped two waves in which the probe beam itself also acts as the read-out beam[29,32–35], and (ii) four-wave mixing (FWM) with a separate writing (probe), read-out (data) beam and reference beam[29,36,37]. For the self-pumped scheme, a relatively rapid response time of 35 ms was achieved but at a relatively low <1-Mbit/s data rate (due to the use of a free-space modulator)[33–35,38]. For the FWM-based approach, a relatively high data rate of 2-Gbit/s was achieved (due to the use of a high-speed fiber-coupled modulator) but the response time of >60 s was too slow to mitigate dynamic turbulence effects[29,37].

In this paper, we achieve both high data rate and rapid response time simultaneously. We experimentally demonstrate the automatic mitigation of dynamic turbulence in an 8-Gbit/s QPSK coherent FSO link using

FWM-based OPC in an undoped GaAs crystal with a <5-ms response time. We transmit a Gaussian probe beam from the Rx to the Tx through emulated turbulence. At the Tx, we create a Gaussian reference beam from the distorted probe beam through single-mode fiber (SMF)-based mode filtering and an optical amplifier. Moreover, we generate a Gaussian beam carrying an 8-Gbit/s QPSK data signal through an SMF-coupled phase modulator. Subsequently, a phase-conjugate data beam is generated through an FWM process in the crystal with the inputs of the probe, reference, and data beams. When the phase-conjugate beam propagates through turbulence to the Rx, turbulence-induced beam distortion and modal coupling are mitigated, and efficient coherent heterodyne detection is enabled. Under emulated dynamic turbulence with a Greenwood frequency of ~260 Hz, our approach shows a ~10-dB reduction in the LO-data mixing power loss and fluctuation. Moreover, our mitigation achieves bit-error rates (BERs) below the 7% forward error correction (FEC) limit for 400 different dynamic turbulence realizations, while a conventional coherent link has ~41% of the realizations above the FEC limit. Compared to prior demonstrations of crystal-based OPC turbulence mitigation, we show ~10,000-fold faster response time for coherent FSO links with >Gbit/s data rate through dynamic turbulence.

## Results

*Matrix representation of turbulence-induced modal coupling for bidirectional beam propagation*

Before showing the concept of our approach, we outline the theory of OPC-based turbulence mitigation by utilizing matrix operations to represent turbulence-induced modal coupling. The wavefront of an optical beam can be distorted when propagating through atmospheric turbulence, causing power coupling from the transmitted spatial mode to other modes[9,12] (e.g., Laguerre-Gaussian ($LG_{\ell,p}$) modes with indices $\ell$ and $p$[39]). This modal coupling process can be approximately represented as matrix manipulation[40]:

$$\boldsymbol{E_{out}} = U\boldsymbol{E_{in}} \qquad (1)$$

where the vectors $\boldsymbol{E_{in}}$ and $\boldsymbol{E_{out}}$ describe the complex coefficients of $LG$ mode components of the input and output optical fields, respectively. Here, we consider a bidirectional beam propagation scenario between a pair of Tx and Rx. Each of $\boldsymbol{E_{in}}$ and $\boldsymbol{E_{out}}$ has $2N$ elements, where the first $N$ elements describe the coefficients of modes at the Tx, and the second $N$ elements describe the modes at the Rx. Here, $N$ corresponds to the number of $LG$ modes that can be transmitted/detected at the Tx/Rx[40]. These $N$ modes can be ranked in the vector by their mode-group orders (i.e., $2p + |\ell|$)[41]. The first element in the vector represents the fundamental Gaussian mode ($LG_{\ell=0,p=0}$). The modal-coupling matrix $U$ with $2N \times 2N$ dimensions can be written as[40]

$$U = \begin{bmatrix} 0 & T_{rt}^{\leftarrow} \\ T_{tr}^{\rightarrow} & 0 \end{bmatrix} \qquad (2)$$

where the $N \times N$ matrix $T_{tr}^{\rightarrow}$ represents the modal coupling for the beam propagating from the Tx to the Rx and $N \times N$ matrix $T_{rt}^{\leftarrow}$ for the beam propagating from the Rx to the Tx. If Tx and Rx can capture all the modes reaching them from the other end, the transmission through turbulence can be considered as a unitary process[24,42]. Therefore, we have

$$T_{tr}^{\rightarrow H} T_{tr}^{\rightarrow} = I \text{ and } T_{rt}^{\leftarrow H} T_{rt}^{\leftarrow} = I \qquad (3)$$

where H denotes the conjugate transpose of a matrix and $I$ is an identical matrix. Moreover, due to the reciprocal property of turbulence[43,44], the matrix $T_{tr}^{\rightarrow}$ is the transpose matrix of $T_{rt}^{\leftarrow}$:

$$T_{rt}^{\leftarrow} = T_{tr}^{\rightarrow T} \qquad (4)$$

Based on Eqs. (3) and (4), we have

$$UU^* = \begin{bmatrix} 0 & T_{rt}^{\leftarrow} \\ T_{tr}^{\rightarrow} & 0 \end{bmatrix} \begin{bmatrix} 0 & T_{rt}^{\leftarrow *} \\ T_{tr}^{\rightarrow *} & 0 \end{bmatrix} = \begin{bmatrix} 0 & T_{tr}^{\rightarrow T} \\ T_{rt}^{\leftarrow T} & 0 \end{bmatrix} \begin{bmatrix} 0 & T_{rt}^{\leftarrow *} \\ T_{tr}^{\rightarrow *} & 0 \end{bmatrix} = I \qquad (5)$$

Eq. (5) shows that the matrix multiplication of $U$ and its complex conjugate $U^*$ results in an identity matrix, which is a key theoretical foundation that supports the use of phase conjugation for turbulence mitigation.

*Turbulence mitigation using OPC for coherent FSO communication links*

As shown in Fig. 1 (a), atmospheric turbulence can significantly degrade the performance of a coherent FSO link. At the Tx, a fundamental Gaussian beam (i.e., $LG_{0,0}$ mode) carrying a data signal ($S(t)$) is transmitted through turbulence to the Rx. The turbulence can cause power coupling from the $LG_{0,0}$ mode to many other modes. According to Eqs. (1) and (2), the turbulence-distorted data beam can be represented as

$$\boldsymbol{E}_{d_r} = U \boldsymbol{E}_{d_t} = \begin{bmatrix} 0 & T_{rt}^{\leftarrow} \\ T_{tr}^{\rightarrow} & 0 \end{bmatrix} \begin{bmatrix} \begin{pmatrix} 1 \\ 0 \\ \vdots \\ 0 \end{pmatrix}_{N \times 1} \\ \begin{pmatrix} 0 \\ 0 \\ \vdots \\ 0 \end{pmatrix}_{N \times 1} \end{bmatrix} \cdot S(t) \qquad (6)$$

where $\boldsymbol{E}_{d_t}$ and $\boldsymbol{E}_{d_r}$ are mode representation of the optical data beams at the Tx and the Rx, respectively, and $S(t)$ is the modulated data signal. Therefore, the turbulence-distorted beam at the Rx contains many modes, which correspond to the first column of the matrix $T_{tr}^{\rightarrow}$. At the Rx, a Gaussian-mode ($LG_{0,0}$) local oscillator (LO) is utilized to optoelectrically mix with the received data beam to recover the data signal using coherent detection. However, the turbulence-distorted data beam is a multi-mode beam, and only the power on the $LG_{0,0}$ modal component (i.e., the first element of the first column of $T_{tr}^{\rightarrow}$) of the beam efficiently mixes with the LO. Such mixing loss can happen for both (i) a free-space-coupled detector due to the orthogonality between the higher-order modes and the Gaussian LO[13] and (ii) an SMF-coupled detector where the higher-order modes at most barely couple into fiber[12,45]. Therefore, coherent detection becomes significantly inefficient, which results in a lower quality of the received data signal.

Figure 1 (b) shows the concept of our approach utilizing OPC to mitigate turbulence in a coherent FSO link. We transmit a continuous-wave (CW) Gaussian-mode probe beam ($E_{p_r}$) from the Rx through turbulence to the Tx.

$$\boldsymbol{E}_{p_t} = U\boldsymbol{E}_{p_r} = \begin{bmatrix} 0 & T_{rt}^{\leftarrow} \\ T_{tr}^{\rightarrow} & 0 \end{bmatrix} \begin{bmatrix} \begin{pmatrix} 0 \\ 0 \\ \vdots \\ 0 \end{pmatrix}_{N \times 1} \\ \begin{pmatrix} 1 \\ 0 \\ \vdots \\ 0 \end{pmatrix}_{N \times 1} \end{bmatrix} \quad (7)$$

At the Tx, the turbulence-distorted probe beam ($\boldsymbol{E}_{p_t}$) acts as an input to an optical phase conjugator for generating a beam with phase-conjugated spatial distribution. In our approach, an FWM-based OPC process in a crystal is implemented. A data-modulated optical signal ($S(t)$) is created and acts as a Gaussian input beam to the FWM process. Therefore, a phase-conjugate data beam ($\boldsymbol{E}_{d_t}$) is generated at the Tx.

$$\boldsymbol{E}_{d_t} = \left(\boldsymbol{E}_{p_t}\right)^* \cdot S(t) = U^*\left(\boldsymbol{E}_{p_r}\right)^* \cdot S(t) \quad (8)$$

When the phase-conjugate beam propagates back along the same path through the turbulence, the turbulence-induced modal coupling is automatically mitigated:

$$\boldsymbol{E}_{d_r} = U\boldsymbol{E}_{d_t} = UU^*\left(\boldsymbol{E}_{p_r}\right)^* \cdot S(t) = \left(\boldsymbol{E}_{p_r}\right)^* \cdot S(t) \quad (9)$$

The mitigated data beam ($\boldsymbol{E}_{d_r}$) at the Rx becomes the conjugation of the single-Gaussian-mode CW probe beam modulated by the data signal, $\left(\boldsymbol{E}_{p_r}\right)^* \cdot S(t)$. As a result, the mitigated data beam can be efficiently mixed with the Gaussian LO and achieve higher quality of the received data as compared to the case without mitigation. Here, Eq. (9) is ideal, if (i) the phase conjugation process is fast enough so that the phase-conjugate data beam propagates through the same turbulence and experiences the same modal coupling ($U$) as the probe beam, and (ii) the Tx and Rx apertures can capture all the modes so that the $U$ is a unitary matrix.

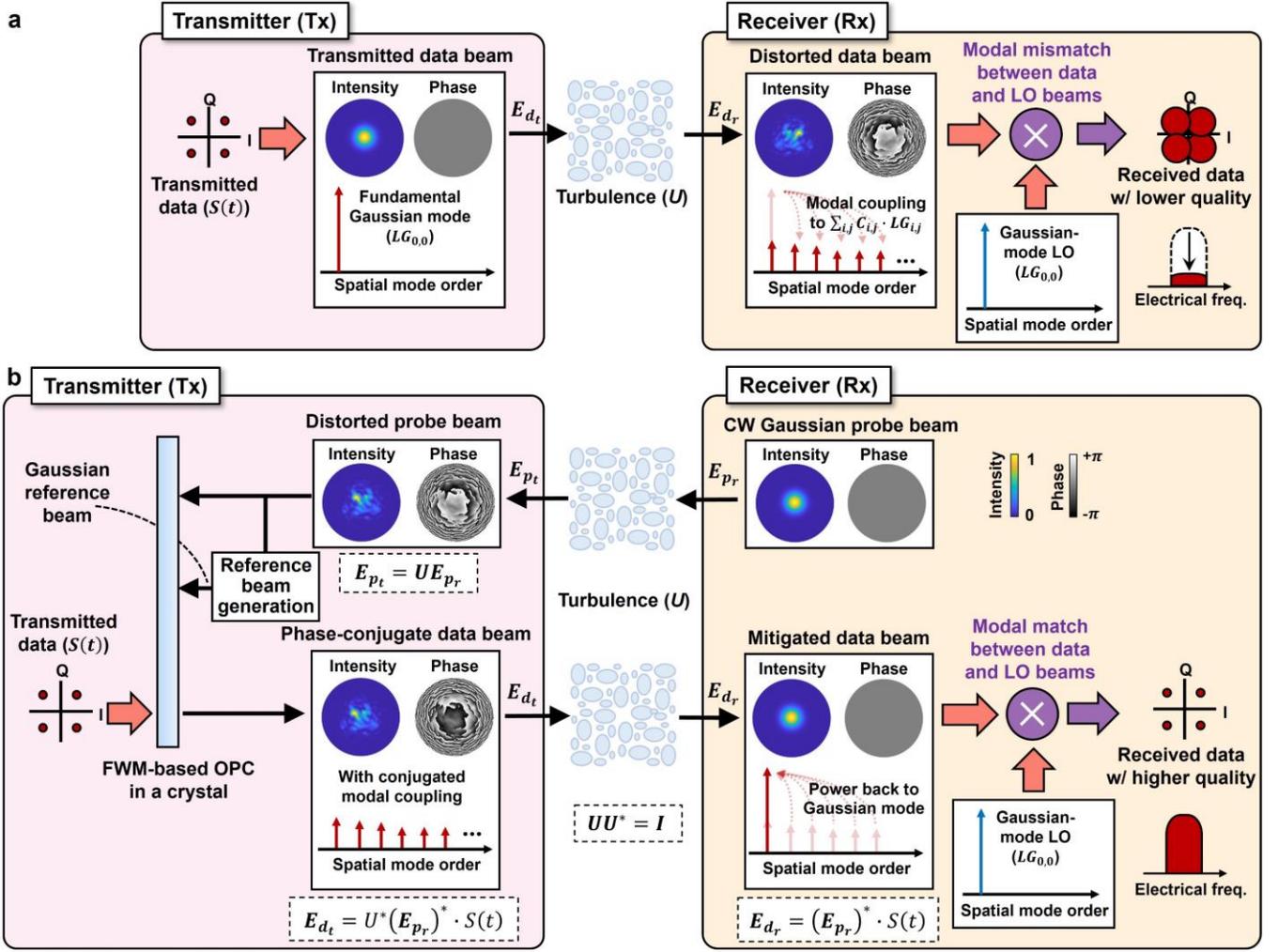

**Figure 1. Concept of utilizing OPC to mitigate turbulence in coherent FSO links.** (a) A conventional coherent FSO link can be significantly degraded by turbulence. The transmitted Gaussian-mode ($LG_{0,0}$) data beam ($E_{d_t}$) carrying data channel ($S(t)$) is distorted by turbulence and its power is coupled to many other modes ($\sum_{i,j} C_{i,j} \cdot LG_{i,j}$). At the Rx, the mixing between the distorted beam and a Gaussian LO beam becomes significantly inefficient in coherent detection due to the modal mismatch between the two beams, thus resulting in lower quality of the recovered data. (b) Our proposed coherent FSO link uses OPC to mitigate turbulence. A CW Gaussian probe beam ($E_{p_r}$) propagates from the Rx to the Tx and is distorted by turbulence. At the Tx, a phase-conjugate data beam is created by a fast (compared with the turbulence dynamics) FWM-based OPC with the inputs of the distorted probe beam ($E_{p_t} = UE_{p_r}$), a Gaussian reference beam, and a Gaussian data channel ($S(t)$). The phase-conjugate data beam ($E_{d_t} = U^* E_{p_r}^* \cdot S(t)$) is transmitted along the reverse path through the same turbulence to the Rx. Due to that the turbulence-induced modal coupling matrix satisfies $UU^* = I$, the data beam ($E_{d_r} = E_{p_r}^* \cdot S(t)$) is recovered to a Gaussian mode and can be efficiently mixed with a Gaussian LO for coherent detection at the Rx.

*Implementation of our phase conjugation architecture based on FWM in a GaAs crystal*

Figure 2 shows a schematic diagram of our architecture of OPC for turbulence mitigation. A CW Gaussian probe beam is generated by Laser 1 at the Rx and propagates to the Tx through dynamic turbulence. At the Tx, the turbulence-distorted probe beam is divided by a beam splitter into two copies. One copy is first coupled into an SMF, then amplified by an optical amplifier with a constant output power, and finally coupled out to free space. In this way of spatial-mode filtering, we generate a collimated Gaussian reference beam that has a fixed power and is mutually coherent with the probe beam. The reference beam and the other copy of the distorted probe beam interfere and illuminate the crystal. Here, we use a GaAs semiconductor crystal, having a fast response time (can be less than 1 ms[46–48]) at near-infrared wavelengths. At the same time, an optical data channel is generated at the Tx by modulating Laser 2 using an SMF-coupled high-bandwidth modulator. This data channel is coupled out to a free-space and collimated Gaussian beam and hits the crystal in a direction opposite to that of the Gaussian reference beam. Through FWM involving the three input beams illuminating the crystal, a phase-conjugate data beam is created and transmitted through the (ideally) same turbulence back to the Rx for coherent detection. The Gaussian LO beam can be created from the same laser source (Laser 1) as the probe beam.

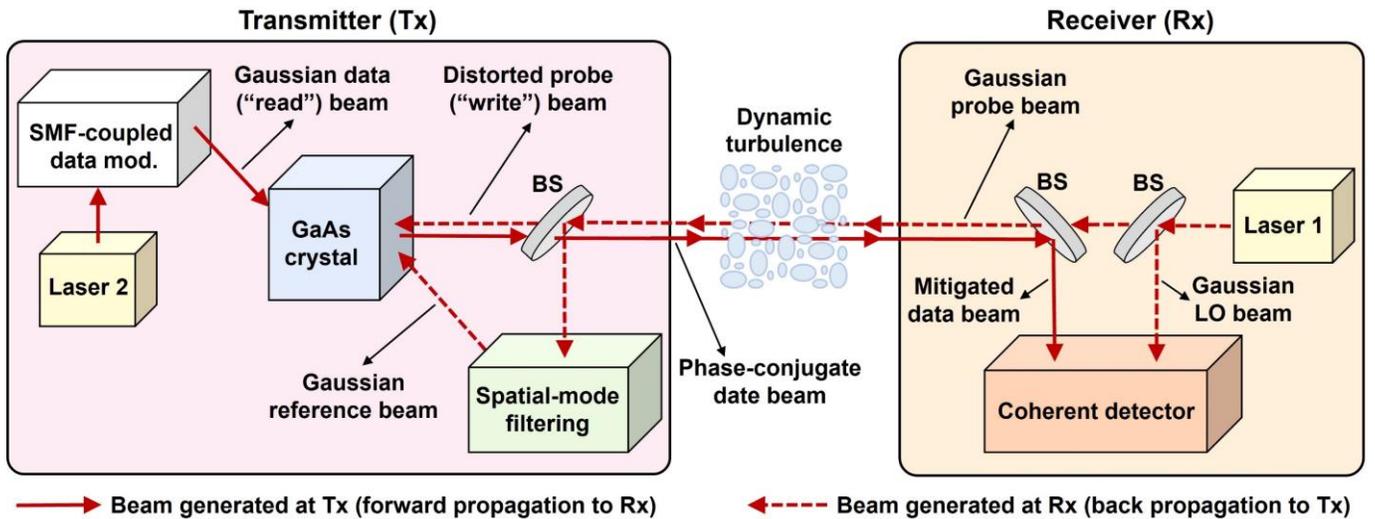

**Figure 2. Our proposed architecture of OPC for turbulence mitigation.** The distorted probe beam is split into two copies. One copy is coupled to an SMF and amplified by an optical amplifier to generate a Gaussian reference beam. An optical data channel is generated by an SMF-coupled modulator and coupled to free space to create a Gaussian data beam. Through four-wave mixing (FWM) in a GaAs crystal, a phase-conjugate data beam can be created and transmitted to the Rx along the reverse propagation direction of the probe beam. Mod.: modulator; BS: beam splitter; SMF: single-mode fiber; Amp.: amplifier.

Such an OPC process can also be explained by a holography model[29]. The interference between the mutually coherent Gaussian reference beam and the distorted probe beam produces interference fringes inside the crystal. Given that the refractive-index change of the crystal is proportional to the light intensity, the intensity of interference creates a holographic grating that is "recorded" inside the crystal[29]. The Gaussian data beam acts as a "read" beam and is diffracted by the grating in a reverse direction of the probe beam.

The diffracted data beam would process the phase conjugation of the spatial information of the probe beam (please see *Supplementary Information Note 1* for additional information).

Here, we compare our phase conjugation architecture with previous demonstrations:

(a) In previous FWM-based OPC architectures[47,49], the coherent reference beam is transmitted from the Rx to the Tx through a path that bypasses the distortion media. The reference beam is transmitted through either a different free-space path or a separate fiber. Such an architecture requires an additional undistorted optical link for the reference beam transmission, a requirement that might not be readily and easily satisfied for FSO communications through turbulence. Our architecture regenerates a coherent Gaussian reference beam from the probe beam without requiring another undistorted link.

(b) The crystal-based OPC has also been achieved in a self-pumped architecture. In such an architecture, the crystal solely "reflects" the distorted probe beam to generate a phase-conjugate beam without the need to involve additional beams in the process[32]. Since the phase-conjugated beam is a multi-mode beam, one needs to use a data modulator that can support many modes, such as a free-space data modulator, to temporally modulate data on the phase-conjugate beam[35]. However, free-space-coupled data modulators generally have much narrower bandwidths than SMF-coupled modulators[50]. Our architecture can use an SMF-coupled modulator, thus enabling higher-rate data transmission.

As a proof-of-concept demonstration, we built the experimental setup as shown in Fig. 3. More details are provided in the *Methods*. At the Rx, a CW beam is generated by a laser at ~1064 nm, amplified by a ytterbium-doped fiber amplifier (YDFA), and coupled out to free space as a Gaussian probe beam. The probe beam propagates from the Rx to the Tx through dynamic turbulence, which is emulated by a rotating phase screen placed around the middle of a ~1-m free-space link. The phase screen is designed based on Kolmogorov turbulence power spectrum statistics[8,12]. The Fried parameter $r_0$ is used to characterize the strength of the emulated turbulence and a smaller $r_0$ means stronger turbulence[9]. Based on the rotation speed (i.e., round/sec.), we can calculate the Greenwood frequency $f_G$ (see *Methods* for the calculation), which is commonly used to characterize the rate of turbulence change[9,10]. At the Tx, we use a beam splitter to create two copies of the probe beam, with one being coupled to an SMF and amplified by a YDFA for generating a Gaussian reference beam. Moreover, we create a Gaussian data beam carrying an 8-Gbit/s QPSK signal by modulating a laser through an SMF-coupled phase modulator. A phase-conjugate data beam is generated by the proposed optical phase conjugator and propagates back toward the Rx through the rotating plate. At the Rx, after being coupled into an SMF, the data beam is mixed with an LO for coherent heterodyne detection. We measure the complex wavefront of the received data beam using off-axis holography and perform *LG* modal decomposition to analyze the turbulence-induced modal coupling[12,51,52].

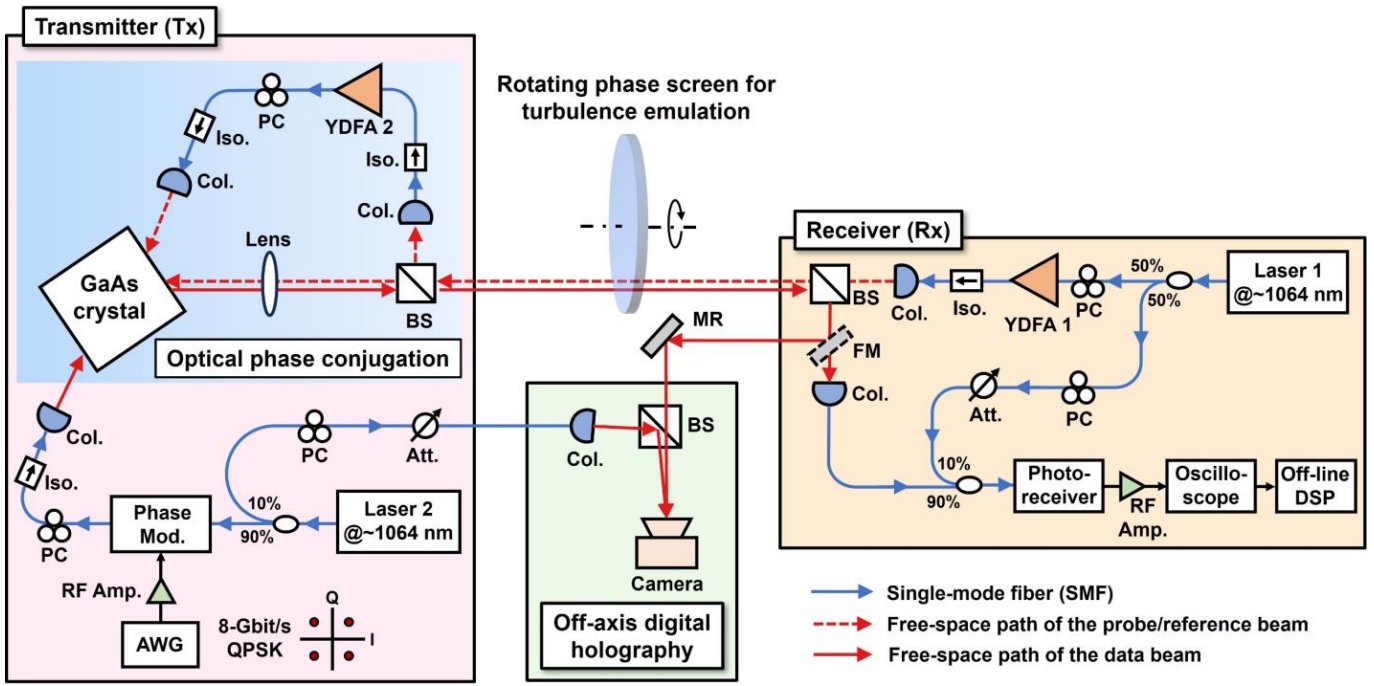

**Figure 3. Experimental setup for turbulence mitigation using OPC in an 8-Gbit/s QPSK coherent FSO link.** At the receiver (Rx), a CW Gaussian probe beam is transmitted through emulated dynamic turbulence to the transmitter (Tx). At the Tx, an 8-Gbit/s QPSK data channel is generated through an SMF-coupled phase modulator and coupled to free space as a Gaussian data beam using a laser with no coherence relation to the probe optical source. A phase-conjugate data beam is created by our phase conjugator and propagates back to the Rx through turbulence. At the Rx, for LG spectrum measurement, the data beams are sent to (by a flip mirror, FM) an off-axis holography setup. For the data detection, the data beam is coupled to an SMF and mixed with a Gaussian-mode LO in a photoreceiver. Heterodyne coherent detection is applied, and offline DSP is used to recover the QPSK signal. AWG: arbitrary waveform generator; Mod.: modulator; Amp.: amplifier; YDFA: ytterbium-doped fiber amplifier; PC: polarization controller; Iso.: isolator; Col.: collimator; BS: beam splitter; MR: mirror; FM: flip mirror; SMF: single-mode fiber; Att.: attenuator; DSP: digital signal processing.

*Modal coupling measurements under dynamic turbulence*

Figure 4 shows the beam profile and modal-coupling measurements of the received data beam without and with phase conjugation mitigation. For the case without mitigation, we transmit the Gaussian data beam from the Tx to the Rx through the same turbulence phase screen along the same path as the phase-conjugate data beam. As shown in Figs. 4 (a) and (b), when we remove the turbulence phase screen, most power of the received beam is on the $LG_{0,0}$ mode for both cases of with and without mitigation. Figures 4 (c) and (d) show the results when the turbulence screen (Fried parameter $r_0$=0.6 mm) is rotating at a speed of 2 round/s, which corresponds to a Greenwood frequency $f_G$ of ~260 Hz. Limited by the frame rate (200 frames/sec) of the camera we used, the profile and modal-coupling measurements are taken every 5 ms. Here, we show 5 successive measurements as examples for each case. A movie containing 100 successive frames is provided in the *Supplementary Video*. As shown in Fig. 4 (c), the intensity/phase profiles are distorted, and the power of the beam is coupled to multiple modes without mitigation. Both distortion and coupling measures are dynamically changing. With mitigation, the intensity/phase profiles of the received beam can be recovered back to Gaussian-like distribution. Moreover, the power of the beam mostly remains on the $LG_{0,0}$ modal component and only coupled to several neighboring modes, which indicates that our approach can effectively mitigate the turbulence-induced modal coupling within a <5-ms response time.

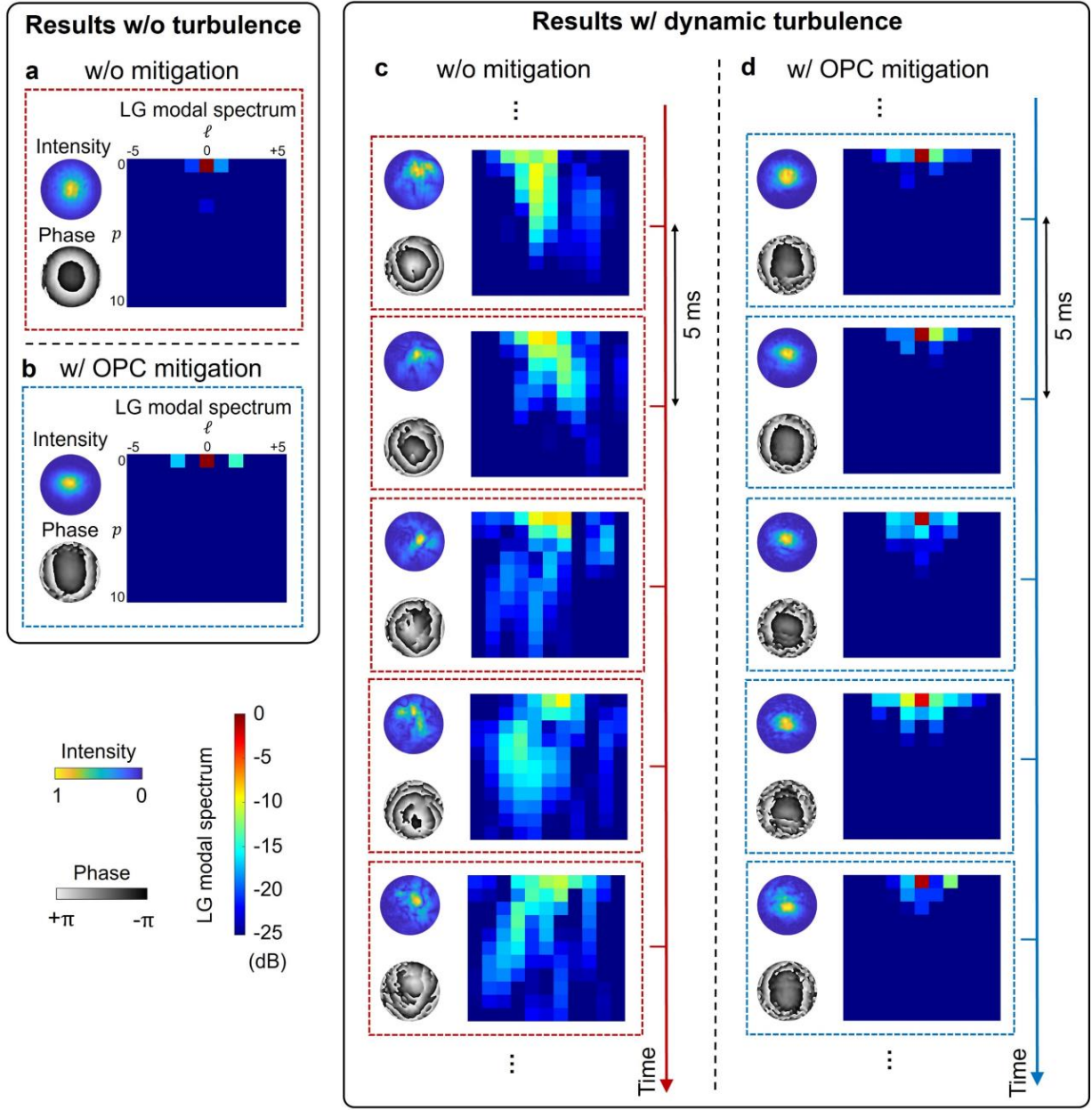

**Figure 4. Experimental results of beam profiles and modal coupling for the received data beam under dynamic turbulence without and with OPC mitigation.** Results under no turbulence (a) without mitigation and (b) with mitigation. Results under dynamic turbulence when the turbulence phase screen ($r_0$ =0.6 mm) is rotating at 2 round/sec (c) without mitigation and (d) with mitigation. For these measurements, we use a camera with a frame rate of 200 frame/sec., corresponding to 5 ms per measurement.

We also measure turbulence mitigation performance using a faster measurement approach. Specifically, we couple the received beam into an SMF and measured the power using a 10 kHz-bandwidth photodiode with a sampling rate of 900k sample/s. This measurement can help to show the fluctuation of the power remaining on the fundamental Gaussian mode under dynamic turbulence. Figures 5 (a-c) show the results also for the case when the turbulence phase screen rotates at a speed of 2 round/s. Based on the measured turbulence-induced SMF-coupling loss, our mitigation approach can reduce the power fluctuation range by ~10 dB. Moreover, we also calculate the power density spectrum in Fig. 5 (c). The frequency components below ~200 Hz are effectively reduced with mitigation, which also indicates that coupled power into the SMF has less fluctuation. In *Supplementary Information Note 2*, we provide the power fluctuation, probability distribution, and density spectrum of the SMF-coupling loss for other phase screen rotation speeds at 0.2, 0.4, 1, and 4 round/sec. As calculated in Fig. 5 (d), the corresponding Greenwood frequency ranges from ~25 Hz to ~500 Hz for different rotation speeds. Figure 5 (e) shows the mean and standard deviation of the SMF-coupling loss, both of which are reduced with our OPC mitigation. However, the reduction tends to be lower when the turbulence phase screen is rotating faster. This might be due to that a faster rotation of the screen results in a larger amount of higher-frequency turbulence changes[53], and the crystal becomes less efficient for mitigating that higher-frequency components.

We also measure the mean and standard deviation of the SMF-coupling loss under weaker turbulence with a larger $r_0$=1.8 mm (see *Supplementary Information Note 3*). With the same rotation speed of the screen, the screen with a larger $r_0$ has a smaller Greenwood frequency $f_G$[9,54]. The results show that the mean and standard deviation under this turbulence can also be reduced by our approach. We note that the results under dynamic turbulence with and without mitigation are measured at separate times due to the limitations of our measurement setup. Therefore, it is difficult to directly compare the two results for the same time labels.

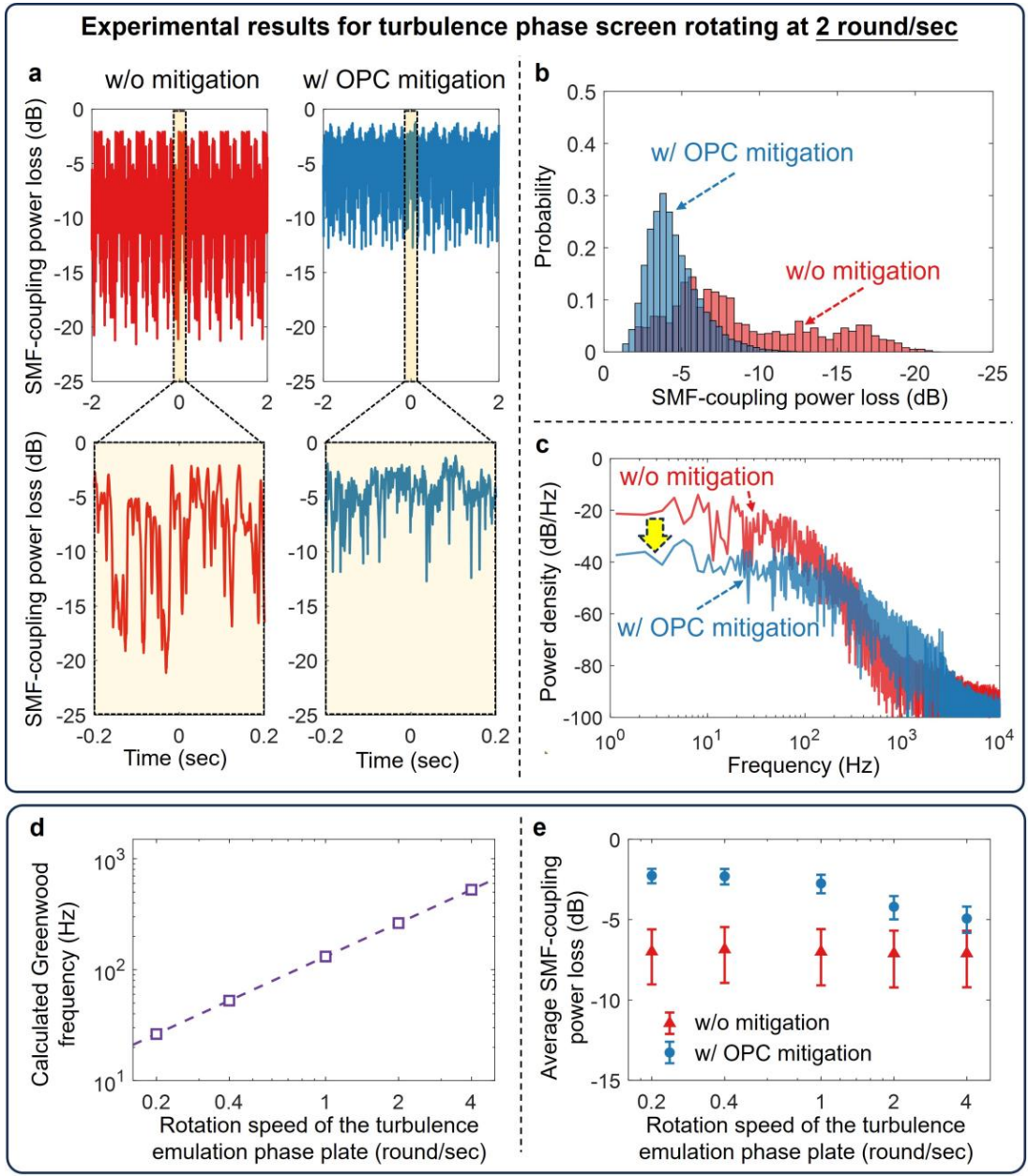

**Figure 5. Experimental results of SMF-coupling power loss of the received data beam under dynamic turbulence.** (a) Power fluctuation, (b) probability distribution, and (c) density spectrum of the SMF-coupling loss without and with phase conjugation mitigation. The results in (a-c) are for the turbulence phase screen ($r_0$=0.6 mm) rotating at 2 round/sec. The power loss is measured by a fiber-coupled photodiode with a 10-kHz bandwidth at a sampling rate of 900k sample/sec. (d) Calculated Greenwood frequency (see *Methods* for the calculation) for different rotation speeds of the turbulence phase screen. (e) Average SMF-coupling loss for different rotation speeds of the turbulence phase screen. The error bars show the standard deviation of the measurements.

*Mitigation of dynamic turbulence in an 8-Gbit/s QPSK coherent FSO link*

Next, we demonstrate the proposed turbulence mitigation in an 8-Gbit/s QPSK FSO communication link. Figure 6 shows the data transmission performance under 400 different turbulence realizations. The results are measured when the turbulence phase screen is continuously rotating at a speed of 2 round/sec corresponding to the Greenwood frequency of ~260 Hz (Fried parameter $r_0$=0.6 mm). To focus on the turbulence mitigation performance, we ensure similar optical powers for the transmitted Gaussian (without mitigation) and phase conjugate data beam (with mitigation). Therefore, they have similar performance with error vector magnitudes (EVMs)[55] at ~18% under no turbulence, as shown in Figs. 6 (a1) and (b1). When there is turbulence, the EVM performance can be degraded up to ~80% without mitigation, while the EVMs with mitigation are below ~35% for all realizations. Moreover, our mitigation can achieve BER values below the 7% FEC limit for all realizations. However, since turbulence can cause strong modal-coupling-induced loss, the performance without mitigation does not achieve the 7% forward error correction limit for ~41% of the realizations.

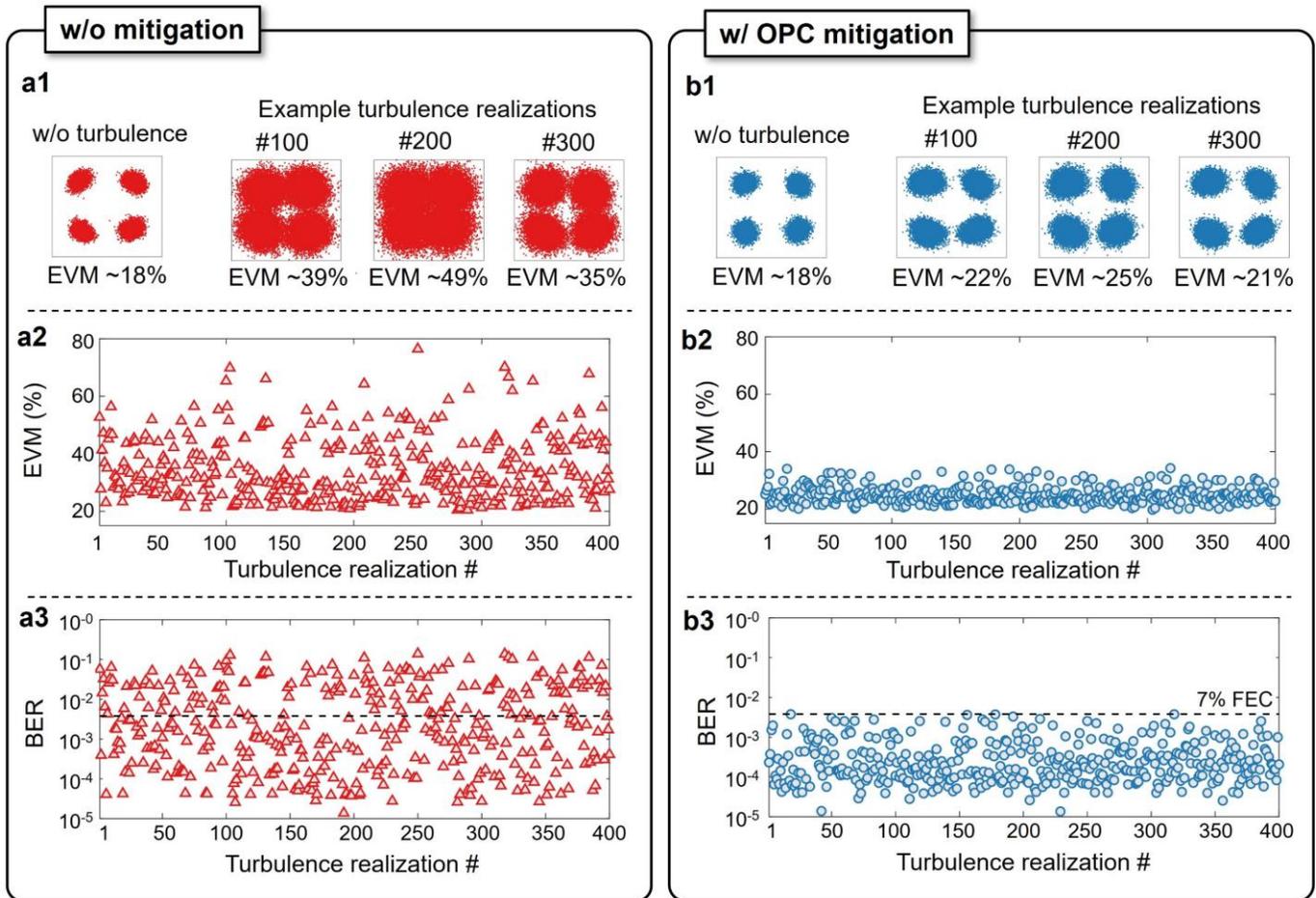

**Figure 6. Experimental results for 8-Gbit/s QPSK data transmission in a coherent FSO link under dynamic turbulence without mitigation and with OPC mitigation.** Results are measured for 400 different turbulence realizations when the turbulence phase screen rotates at 2 round/sec., including data constellations in (a1) and (b1), EVM performance in (a2) and (b2), and BER performance in (a3) and (b3). Note that we measure the results with and without mitigation at separate times due to the limitations of our measurement setup. Therefore, the results with the same realization label may correspond to different turbulence realizations and are difficult to be directly compared.

## Discussion

In this paper, we show turbulence mitigation at a wavelength of ~1064 nm, dictated by the efficient working wavelength of the GaAs crystal. The data rate and modulation format demonstrated in our experiment are limited by our current modulation equipment at ~1064 nm. We believe our approach is a general system architecture that can potentially be applied to other wavelengths and supports higher data rates with more complex formats (e.g., QAM), given other appropriate crystals and high-performance devices (e.g., I/Q modulator) can be used.

In our OPC architecture, a Gaussian reference beam is regenerated from the turbulence-distorted CW probe beam by SMF coupling and YDFA amplification. We used a high-power YDFA in our experiment having a minimum input optical power of ~-3 dBm. When the turbulence is stronger, the probe power coupled to the YDFA tends to be weaker and might be lower than the required input power. Therefore, our YDFA might not be able to amplify such weak power for reference beam regeneration. We note that one potential way to handle this issue is using optical injection locking that can amplify a CW input light with smaller power[56,57].

For optical nonlinear processes, power efficiency is always a potential concern. The efficiency in our demonstration for the phase conjugation is roughly $5\times10^{-4}$ (conversion between the Gaussian data beam to the phase-conjugate data beam). However, this could be increased by orders of magnitude by optimizing the crystal parameters and geometrical arrangements of the probe and reference beams, as well as applying an external electrical voltage to the crystal[47,58].

# Methods

*Detailed description of the experimental setup:*

At the Rx, we generate a CW probe beam from Laser 1 at ~1064 nm, amplify it by YDFA 1, couple it to free space as a collimated Gaussian beam with a waist diameter of ~2.5 mm, and transmit it from the Rx to the Tx in a ~1-m free-space link. In the middle of the link, we place a rotating round phase screen to emulate dynamic turbulence effects with the Fried parameter $r_0$ of 0.6 mm. When the probe beam propagates through the phase screen to the Tx, its wavefront gets distorted. We use a beam splitter (50:50) to create two copies of the distorted probe beam. One copy is coupled to an SMF and amplified by YDFA 2. The YDFA 2 is working with a fixed output power. After amplification, the beam is coupled out to free space as a collimated Gaussian reference beam. The reference beam and the other copy of the distorted probe beam cross with an angle of ~60º, and interfere inside a GaAs crystal to "record" the turbulence distortion. The GaAs crystal has a dimension of 5×5×5 mm³. We use a lens to reduce the size of the probe beam to ensure good spatial overlap of the two beams[36].

At the Tx, we modulate Laser 2 at ~1064 nm with an SMF-coupled phase modulator to generate an 8-Gbit/s QPSK optical data channel. We couple the data channel out to free space as a Gaussian data beam and transmit it in the opposite direction of the Gaussian reference beam through the crystal. The Gaussian data beam "reads" the turbulence distortion and generates a phase-conjugated data beam that propagates toward the Rx in a reverse direction of the probe beam. The power of the distorted probe beam, the Gaussian reference beam, and the Gaussian data beam illuminated on the crystal are ~100 mW, ~100 mW, and ~20 mW, respectively. The power of the generated phase-conjugate data beam is ~0.01 mW.

When the phase-conjugate beam arrives at the Rx, we use a flip mirror to switch the beam propagation path for different measurements. To analyze wavefront distortions and modal coupling induced by turbulence, we measure the complex wavefront of the received beam and perform *LG* modal decomposition using off-axis holography. For the off-axis holography, some power of Laser 2 is coupled out as a reference beam, and the camera is operated at a frame rate of 200 frames/sec.

To detect the data channel, the phase-conjugated data beam is coupled into an SMF and optoelectrically mixed with an LO in an SMF-coupled photoreceiver. The LO is generated from Laser 1 through a fiber coupler and its power is controlled to be ~0 dBm. Without turbulence, the power of the received phase-conjugate data beam coupled into the SMF is ~-29 dBm. To recover the QPSK data signal, we use a coherent heterodyne detection approach, in which the LO-data mixing term is in an intermediate frequency (IF) in the electrical domain. We control the IF to be ~6 GHz by setting the frequency difference between Laser 1 (LO beam) and Laser 2 (data beam). The data signal is recorded by a real-time digital sampling oscilloscope and processed using offline DSP for data recovery and analysis. In our DSP, the recorded sequence is firstly digital frequency down-converted to the baseband to extract the I and Q channels. Subsequently, the I/Q complex signal is processed and recovered sequentially through (i) channel equalization using constant modulus algorithm (CMA)[59], (ii) frequency-offset estimation using a 4th-power-FFT approach[60], and (iii) phase recovery using blind-phase-searching algorithm[61]. Finally, EVMs and BERs of the data channel are calculated for the evaluation of the performance.

*Calculation of the Greenwood frequency $f_G$ of turbulence:*

The Greenwood frequency, $f_G$, is a measure of the rate at which a beam is affected by turbulence[10]. The $f_G$ can be calculated for a given a wind speed $V$ and Fried parameter $r_0$, by[9]

$$f_G = 0.43 \frac{V}{r_0} \tag{10}$$

For our emulated turbulence, the beam illuminates the phase screen at a location that is $R=3$ cm far from the center of the screen. To emulate dynamic turbulence, the phase screen was made to rotate at different speeds of $n=$ 0.2, 0.4, 1, 2, and 4 round/s. The emulated wind speed is estimated by $V = 2\pi R n$, resulting in the following expression for $f_G$:

$$f_G = 0.43 \frac{2\pi R n}{r_0} \tag{11}$$

**Data availability:** All data, theory details, and simulation details that support the findings of this study are available from the corresponding authors upon reasonable request.

**Competing Interests:** The authors declare no competing interests.

**Acknowledgments**

This work is supported by the Office of Naval Research through a MURI award N00014-20-1-2558 and the Airbus Institute for Engineering Research.

# Supplementary Information: Automatic Mitigation of Dynamic Turbulence Using Optical Phase Conjugation for Coherent Free-Space Optical Communications

Huibin Zhou[1,*], Xinzhou Su[1], Yuxiang Duan[1], Yue Zuo[1], Zile Jiang[1], Muralekrishnan Ramakrishnan[1], Jan Tepper[2], Volker Ziegler[2], Robert W. Boyd[3], Moshe Tur[4], and Alan E. Willner[1,*]

1. Department of Electrical and Computer Engineering, Univ. of Southern California, Los Angeles, CA 90089, USA
2. Central Research and Technology, Airbus, Taufkirchen, Germany
3. Institute of Optics, University of Rochester, Rochester, New York 14627, USA
4. School of Electrical Engineering, Tel Aviv University, Ramat Aviv 69978, Israel

Corresponding emails*: H.Z. (huibinzh@usc.edu) or A.E.W. (willner@usc.edu)

**Note 1: More explanations on the holography model for the crystal-based phase conjugation**

Here, we provide a more detailed explanation for the concept of phase conjugation using the holography model. As shown in Fig. S1, the reference beam and the probe beam interfere inside the crystal and form a holographic grating. We can view the transmission function of the holographic grating ($T$) to be related to the interference-induced refractive-index changes[1].

$$T \propto (A_{ref} + A_p)(A_{ref} + A_p)^* = |A_{ref}|^2 + |A_p|^2 + A_{ref}^* A_p + A_{ref} A_p^* \quad (S1)$$

where $A_{ref}$ and $A_p$ are spatial complex envelopes of the reference and probe beams, respectively. Moreover, since the Gaussian reference beam and Gaussian data beam are two collimated counter-propagating beams, the complex amplitude $A_{data\_G}$ of the Gaussian data beam satisfies[1]:

$$A_{data\_G} \approx A_{ref}^* \quad (S2)$$

According to Eqs. (S1) and (S2), the transmission field $A_{data\_G}'$ of the Gaussian data beam after propagating through the grating can be expressed as

$$A_{data\_G}' \propto T A_{data\_G} \approx T A_{ref}^* = \{|A_{ref}|^2 + |A_p|^2\} A_{data\_G} + A_p (A_{ref}^*)^2 + A_{ref} A_{data\_G} A_p^* \quad (S3)$$

The third term on the right side of the Eq. (S3) corresponds to the phase conjugate data beam $A_{data\_PC}$ and can be written as[1]

$$A_{data\_PC} \propto A_{ref} A_{data\_G} A_p^*. \quad (S4)$$

Thus, the phase conjugate data beam has a conjugate spatial distribution of the probe beam.

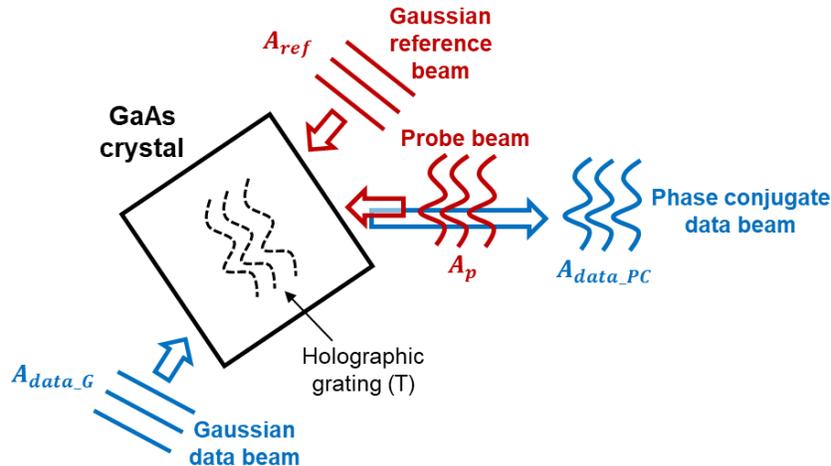

Figure S1. Schematic diagram of the holography modal for the optical phase conjugation in a GaAs crystal.

**Note 2: Power fluctuation, probability distribution, and density spectrum of the SMF-coupling loss for the phase screen rotating at 0.2, 0.4, 1, and 4 round/sec.**

In the Main text of the manuscript, we show the power fluctuations, probability distribution, and density spectrum of the SMF-coupling loss for a phase screen ($r_0$=0.6 mm) rotation speed of 2 round/sec. Here, we provide the corresponding results for phase screen rotation speed at 0.2, 0.4, 1, and 4 round/sec (Figs. S2-S5). For the cases without mitigation, the probability distribution of the loss remains almost the same for different rotation speeds. However, the density spectrum extends to a higher frequency for a faster rotation speed, which means that faster dynamic turbulence effects exist. With mitigation, the fluctuation of SMF-coupling loss can be suppressed for different rotation speeds. However, the suppression is less effective for faster rotation speeds, which correspond to the calculated average and standard deviation shown in Fig. 5(e) of the Main text.

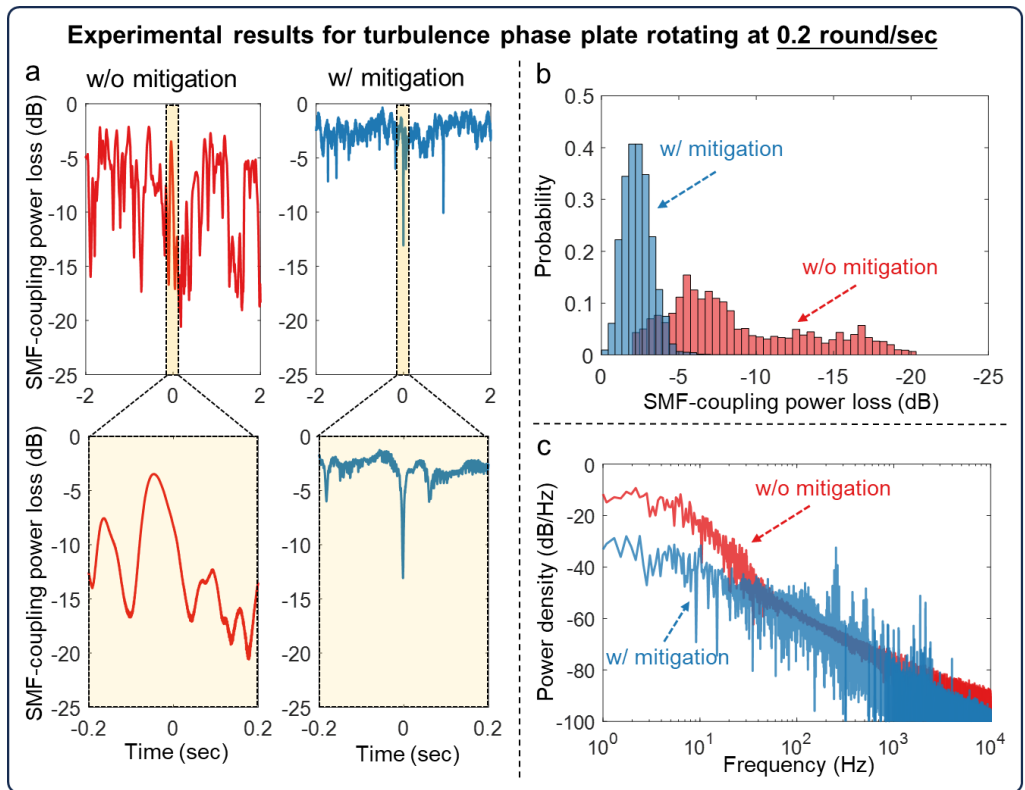

Figure S2. Experimentally measured (a) power fluctuation, (b) probability distribution, and (c) power spectral density of the SMF-coupling loss for the turbulence phase screen rotation speed at 0.2 round/sec.

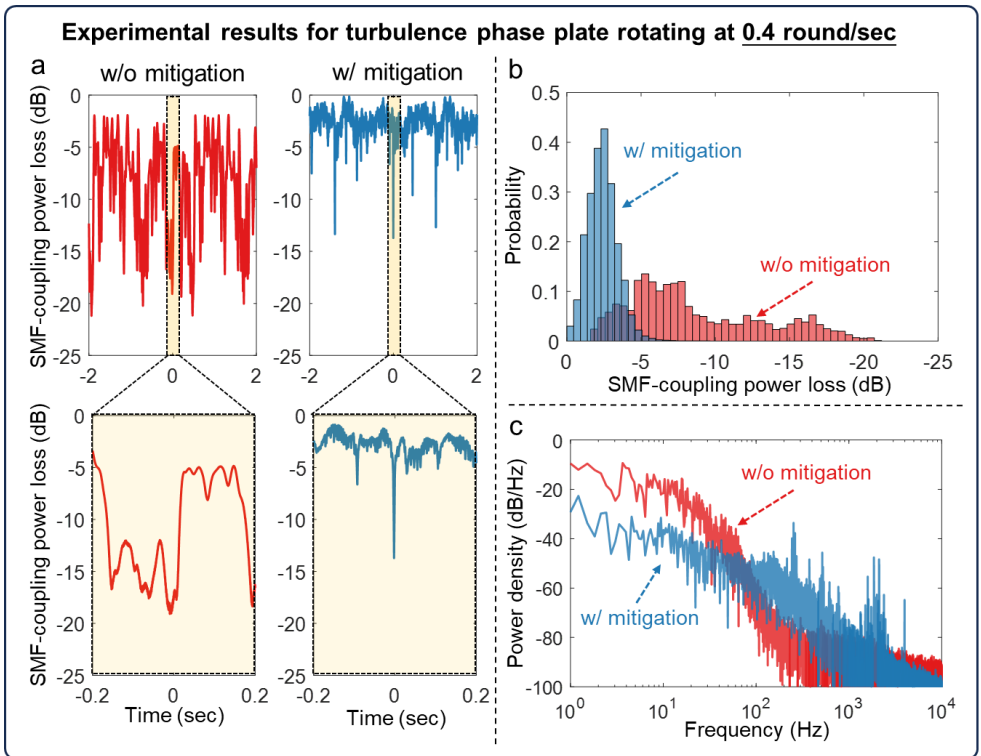

Figure S3. Experimentally measured (a) power fluctuation, (b) probability distribution, and (c) power spectral density the SMF-coupling loss for the turbulence phase screen rotation speed at 0.4 round/sec.

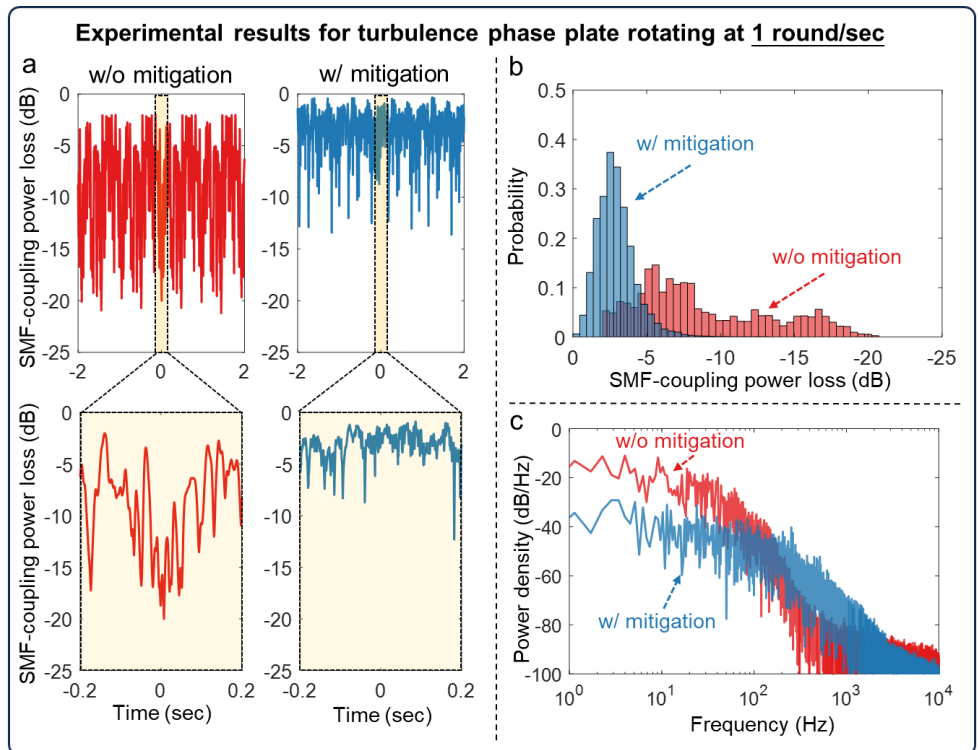

Figure S4. Experimentally measured (a) power fluctuation, (b) probability distribution, and (c) power spectral density of the SMF-coupling loss for the turbulence phase screen rotation speed at 1 round/sec.

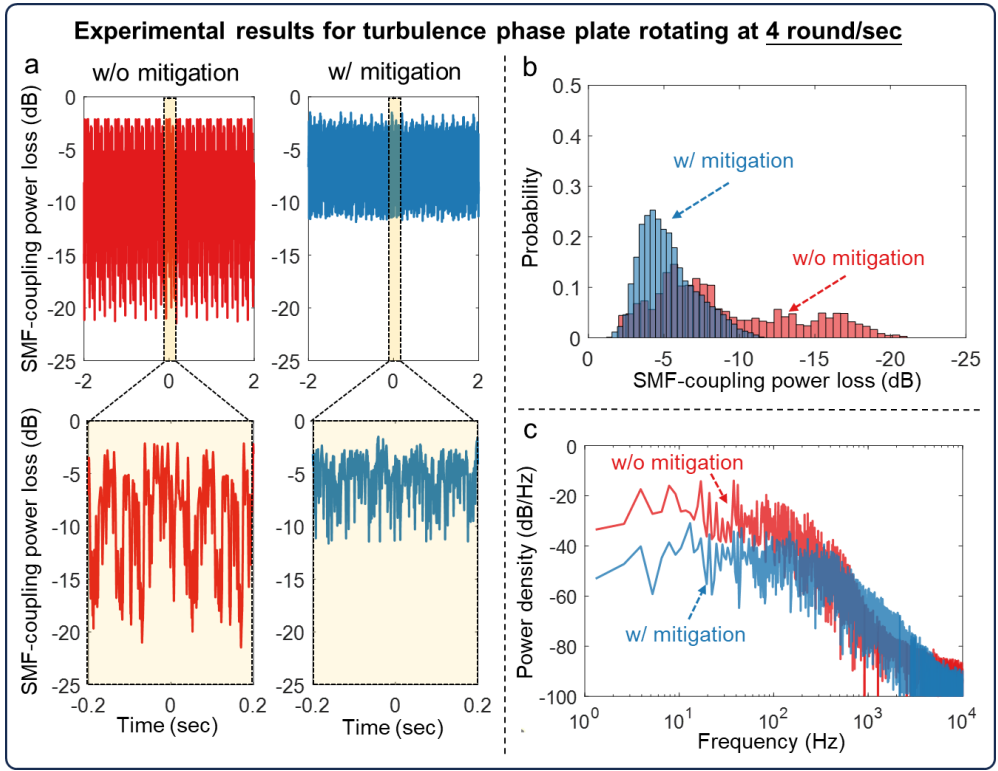

Figure S5. Experimentally measured (a) power fluctuation, (b) probability distribution, and (c) power spectral density of the SMF-coupling loss for the turbulence phase screen rotation speed at 4 round/sec.

**Note 3: Measurement of the mean and standard deviation of the SMF-coupling loss for a turbulence phase screen with Fried parameter $r_0$=1.8 mm.**

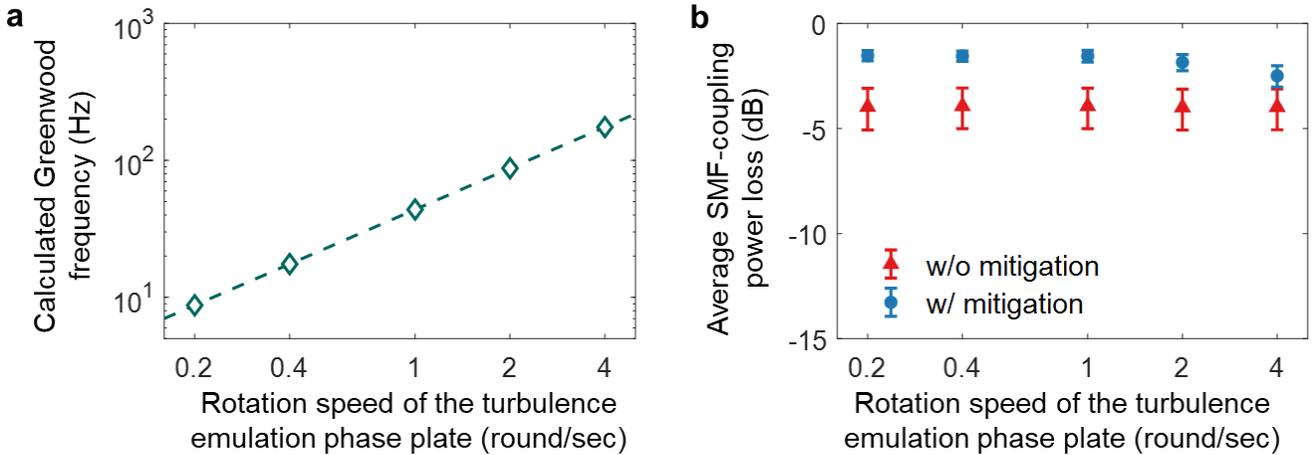

Figure S6. (a) Calculated Greenwood frequency for different rotation speeds of the turbulence phase screen (Fried parameter $r_0$=1.8 mm). (b) Average SMF-coupling loss for different rotation speeds of the turbulence phase screen. The error bars show the standard deviation of the measurements.

In the Main text of the manuscript, we show the mean and standard deviation of the SMF-coupling loss with a Fried parameter $r_0$=0.6 mm. Here, we show the results for weaker turbulence using a phase screen emulating a larger Fied parameter of $r_0$=1.8 mm. Fig. S6(a) is the calculated Greenwood frequency $f_G$. Since $f_G$ is inversely proportional to $r_0^{\,2-4}$, the calculated $f_G$ here is smaller than in Fig. 5(d) of the Main text. As shown in Fig. S6(b), the mean and standard deviation of the average power loss are again reduced by our mitigation approach. Moreover, the SMF-coupling loss is smaller for such weaker turbulence, compared to the results of Fig. 5(e) (Main text), drawn for 3-times stronger turbulence.